# Mindsets and Management:
# AI and Gender (In)Equitable Access to Finance

Genevieve Smith
genevieve.smith@kellogg.ox.ac.uk, University of Oxford | University of California, Berkeley

***Note: This paper has been accepted for presentation at [ACM FAccT 2025](#).***

## ABSTRACT

Various companies are developing apps that collect mobile phone data and use machine learning (ML) to provide credit scores - and subsequently, opportunities to access loans - to groups left out of traditional banking. This paper draws on interview data with leaders, investors, and data scientists at fintech companies developing ML-based alternative lending apps in low- and middle-income countries (LMICs) to answer the question: *In what ways do the underlying logics, design choices, and management decisions of ML-based alternative lending tools by fintechs embed or challenge gender biases, and how do these choices influence gender equity in access to finance?* Findings reveal developers follow "gender blind" approaches, grounded in beliefs that ML is objective and data reflects the truth. This leads to a lack of grappling with the ways data, features for creditworthiness, and access to apps are gendered. Overall, tools increase access to finance, but not gender equitably: Interviewees report less women access loans and receive lower loan amounts than men, despite women being better repayers. Fintechs identify demand- and supply-side reasons for gender differences, but frame them as outside their responsibility. However, that women are observed as better repayers reveals a market inefficiency and potential discriminatory effect, which can be further linked to profit optimization objectives. This research introduces the concept of *encoded gender norms*, whereby without explicit attention to the gendered nature of data and algorithmic design, AI technologies reproduce existing inequalities. In doing so, they reinforce gender norms as self-fulfilling prophecies. The idea that AI technology is inherently objective and, when left alone, "fair", is seductive and misleading. In reality, algorithms reflect the perspectives, priorities, and values of the people and institutions that design them.

## 1 INTRODUCTION

In 2012, John[1] became obsessed with a digital lending side project. His goal was to help unbanked entrepreneurs access finance by using machine learning (ML) to assess alternative data on peoples' smartphones – which were proliferating rapidly in low-and middle-income countries (LMICs) – to provide a credit score. As an American man living and working in rural Kenya in his mid-20s, he volunteered at a microfinance charity and recognized the transformative power of capital for micro-entrepreneurs, many of whom remained stuck in poverty. He also witnessed the barriers that kept poor, trustworthy borrowers outside of formal financial systems, despite their hard work. This realization inspired his career trajectory and mission to unlock capital for those excluded from financial opportunities. Since launching his business nearly a decade ago, it has reached millions of people in Kenya and beyond. As this and other similar algorithmic lending apps scale in the Global South, it raises critical questions about how the use of ML in credit assessment can enhance financial inclusion and reinforce or mitigate gender biases in access to finance.

This paper – drawing on a theoretical basis from feminist STS – examines the underlying logics, design choices, and management decisions of ML-based credit assessment tools by fintechs to explore if and how they embed or challenge gender biases, and the impact these systems have on gender equity in access to finance. It is informed by qualitative interview data with corporate leaders, investors/funders and data scientists at fintech companies developing and managing ML-based alternative lending apps. By investigating these dynamics, this paper seeks to contribute to ongoing debates about the role of artificial intelligence (AI) in shaping inclusive financial systems, as well as the interaction between algorithmic decision making and equity more broadly. Findings also have practical implications for fintechs, policymakers, and funders.

Findings are presented in two sections. The first examines the underlying American banking logics that inform the definition of "creditworthiness" and how this translates into algorithmic choices, alongside Silicon Valley logics that affirm machine learning as objective and data as truthful. In the second, I explore how these logics inform design choices related to building "gender blind" algorithms that lead to a lack of grappling with the ways data, features or proxies for creditworthiness, and access to apps are gendered and reflect inequities. I outline fintechs' perceptions regarding why women receive less loans and at lower amounts than men before providing socio-structural explanations to expose how gendered inequities are reproduced within fintech technologies and through lending decision-making. I conclude with a discussion on my findings, including highlighting an *objective algorithm paradox*, in which the belief in ML's objectivity results in being 'blind' to certain demographics in ways that hide the gendered nature of algorithms and perpetuate discrimination. I expand on existing theories of algorithmic bias and gender equity by introducing the concept of *encoded gender norms* and exploring its relevance in the context of ML-based credit assessment and algorithmic decision-making.

[1] All names of interviews are pseudonyms.

## 2 BACKGROUND

### 2.1 Machine learning unlocks new opportunities for financial inclusion

Development theories emphasize the key role that access to finance can play in mitigating inequality and poverty. Financial market imperfections and lack of access to finance are seen as critical reasons for persistent income inequality and reduced economic growth [4]. Tackling financial market imperfections and enhancing financial inclusion can accelerate economic growth, as well as reduce income inequality and poverty; while access to finance for firms can promote entrepreneurship and innovation, resulting in firm growth and broader economic gains [4]. There are 1.4 billion adults considered “unbanked” and outside of traditional financial folds [51].

Despite large recent gains in financial access, persistent divides remain along axes of gender, socio-economic status, education and more. Women’s account ownership is 6 percentage points lower than men’s in LMICs [51]. Linked to persistent gender discrimination and limiting gender norms, women – and those with lower socio-economic status – are more likely to lack identification, lack a mobile phone, live further from formal financial services, and need more support to use financial accounts [51]. Women also face issues accessing formal finance due to lack of collateral, lack of credit history, and gender discrimination from loan officers [17].

Mobile phones are increasingly key in advancing opportunities for digital payments, savings, and borrowing – a trend catalyzed from COVID-19. Worldwide financial account ownership has reached 76% of people (71% in LMICs) with mobile phones playing a key role fueling growth in Sub-Saharan Africa and for women particularly [51]. Mobile phones and alternative data found on them – when combined with ML – open new opportunities to facilitate loans to those otherwise unable to access them. ML-based alternative lending tools collect mobile phone data and utilize ML to provide credit scores – and subsequent access to loans – to those left out of traditional banking folds in LMICs. These smartphone apps ask for permission to view data on a user’s smartphone and collect real-time data [7]. The application can have access to various data stored on the device [42]. The ML model assesses such data to make predictions about one’s creditworthiness and facilitate access to loans, which tend to be small with high interest rates and short repayment windows.

Use of and investment in fintech for alternative lending is skyrocketing globally alongside excitement over its economic and social impact potential. The alternative financing market was valued at $US10.82 billion in 2022 [26]. Meanwhile the global AI in credit scoring market is set to experience significant growth with a projected compound annual growth rate of 25.9% from 2024 to 2031 [30]. India and Kenya are two of the most common markets for these apps. In Kenya, 77% of borrowers have taken only digital loans, and in 2018, over 90% of loans taken were digital [38]. Two of the largest fintech firms with ML-based alternative lending apps in LMICs (Branch and Tala, both headquartered in California) have over 4 million and 6 million users, respectively, with US$3.3 billion disbursed to date between the two firms [43]. Meanwhile, multilateral organizations, NGOs, and government agencies are eager to support and fund these types of “AI for good” systems. Despite their popularity, research on impacts is lacking.

## 2.2 **Historical roots of "creditworthiness" & credit scoring**

ML-based credit assessment tools are some of the newest mechanisms for evaluating the age-old question of whether someone will repay a loan. The concept of "creditworthiness" as it is popularly applied and understood today, is rooted in American history, where it evolved alongside ideals of meritocracy, individualism, and moral judgment. The notion of "creditworthiness" developed in America was tied not simply to wealth and one's familial economic history (as lending had been done in other areas such as the United Kingdom). Rather, creditworthiness was something anyone could hypothetically attain, as it was linked not only to wealth, but one's character and trustworthiness. The concept of creditworthiness in itself is an American phenomena that has roots tied to the American dream, being able to "lift yourself up by your bootstraps", and the idea of meritocracy. It is defined as both the ability and willingness to repay, closely linked to the "3 C's": Capital, capacity, character [35]. Capital included assets, liabilities, property owned by individuals, and assets available. Capacity included one's age, experience in business, past employment, known history of success or failure. Character – which included exploring one's work habits, local reputation and personal life – was centered around the perceived honesty and morality of the borrower. Assessing someone's creditworthiness was seen as more of an "art", with a recognition that subjectivity and personal opinion was part of the credit assessment process [35]. In the 1950s, two pioneers, Bill Fair and Early Isaac, took the "art" of credit assessment to data science when they developed the credit score; thereby making creditworthiness a function of statistical calculation [22]. Creditworthiness became automated and presented as a number.

## 2.3 **Gender bias in AI**

Research has identified pervasive biases – which manifest in datasets, algorithms, and use of AI systems – resulting in discrimination and illustrating that technology is not neutral [5, 21, 39]. Bias is often baked into datasets that ML systems learn from. Data is not objective and reflects pre-existing social and cultural biases [14, 20]. Further, men are seen as the "human default" resulting in ubiquitous gender data gaps [14]. Bias can also result from under-representation of different groups in training datasets, resulting in lower accuracy for individuals from those groups [10]. Within algorithms, selection of proxies can lead to bias – as illustrated through a widely used healthcare algorithm in the US that falsely concluded Black patients as healthier than equally sick White patients when using health costs as a proxy for health needs ignoring that Black Americans spend less money on healthcare [40]. How ML systems are used can also result in discriminatory outcomes – e.g., if used in a different context or for a different population from which it was originally developed or operationalized [24]. Bias in AI systems can be hard to spot and diagnose. due to their "black box" nature [27].

It matters who develops AI systems. AI systems are created in certain contexts by humans and are classification technologies [50]. The perspectives, knowledge, values, and priorities of those who develop and manage AI systems will be integrated into their design and operation. People inform what AI systems are designed to optimize for and the problem AI is solving, as well as how decisions are made in dataset selection, model development, tool management, and more. The lack of diversity in STEM is widely acknowledged, with those wielding power in tech being disproportionately elite, straight, White, able-bodied, cisgender men from the Global North [20].

### 2.4 Existing research on fintechs & gender in ML-based alternative lending

Existing research on ML-based alternative lending tools in LMICs is thin. It primarily assesses how alternative data and ML-based lending tools can (a) more accurately predict loan repayment for underbanked people [2, 6, 7, 37] and/or (b) mitigate bias found in formal credit assessment mechanisms [3]. Some research examines welfare impacts resulting from access to loans through ML alternative lending tools, including positive impacts and harms related to default and debt traps [7, 8, 43, 47]. Overall, that greater access to finance from these tools results in net positive impacts for individuals alongside broader economic growth and poverty reduction is not a given. There is no research on those deemed *not* creditworthy and bias herein, as well as how those impacts link to the ways tools are conceptualized, designed, and managed. While research on gender biases is lacking, a review of three studies on fintech firms offering digital credit finds that men are disproportionately self-selecting as potential borrowers [43].

## 3 METHODS

The findings in this paper are drawn from 25 semi-structured interviews conducted with corporate leaders, data scientists, and investors at fintech companies developing and managing ML-based alternative lending apps in LMICs. I utilized purposive sampling to identify and recruit the study sample, which followed a landscape review of ML-based alternative lending companies. To recruit, I sent emails or messages through LinkedIn with the interview request. I also attended events whereby target interviewees were speaking or attending. I also utilized snowball sampling. Over 160 participants were invited for interviews and 25 interviews were conducted between August 2023 and February 2024. Each interview lasted approximately one hour, allowing for in-depth discussions with participants. The interviews were organized around a Topic Guide that delved into conceptualization, design, and management of the AI tools. Interviews were transcribed using a speech-to-text transcription service. One interviewee did not consent to their interview being recorded or to quotes being used. In analyzing the data, I utilized an abductive approach through combining deductive components building from my theoretical basis and analytical framework, as well as inductive using a grounded theory approach to allow important themes to emerge. I drew on reflexive thematic analysis, subsequently reviewing and refining themes to identify patterns and core elements in the data. NVivo software facilitated the coding process. Drawing on reflexive thematic analysis [9], after familiarization with the data round, I developed an initial generation of codes. I developed and reviewed themes from the codes before subsequently refining my codebook and writing.

Interview participants were largely male (72%). Of the seven female participants (28% of the sample), one was in a leadership data science role, and three were in leadership (non-data science) positions. These numbers are reflective of the fintech industry more broadly and the lack of female representation, particularly in leadership and data science roles. Fintech is one of the few sectors that combines two historically male dominated industries: finance and technology [25]. Globally, only 1.5% of fintech companies are solely founded by women [23]. Of the fintech companies interviewed, only one fintech was solely founded by a woman (although I did not interview her, but rather another leader at the company). Overall, my sample reflects the lack of gender diversity in the fintech industry.

This paper draws on a theoretical framework integrating insights from Science and Technology Studies (STS), feminist theory, and postcolonial theory that emphasize how technology development is not neutral, but rather shaped by individual choice and power dynamics [34, 36, 50]. Specifically, I use Donna Haraway's "situated knowledge" and Ruha Benjamin's concept of "default discrimination" to analyze how machine learning-based credit assessments in LMICs embed or challenge gendered power structures. Haraway's "situated knowledge" posits that all knowledge is produced from specific, socially, and historically situated standpoints [28]. I apply this to critically assess design choices that reflect the perspectives and priorities of their creators, which are shaped by the cultural and institutional logics. "Technofeminism" supports this analysis by highlighting how the social context of technology developers influences the design choices embedded in these tools, allowing me to critique and unpack choices by algorithmic developers and managers [49]. Benjamin's concept of "default discrimination" informs the gap between intention and outcome in technology design. It critiques how even well-intentioned algorithms, designed with the aim of fairness or inclusion, can perpetuate existing biases if they fail to address the underlying structural inequalities embedded in the data and development processes [5].

## 4 FINDINGS A: UNDERLYING LOGICS & MINDSETS

Using machine learning to assess the credit of unbanked and underbanked people in LMICs emerged in the last 15 years. Several of the leading and largest fintechs in this space have roots in the Bay Area and Silicon Valley. Two primary mindsets and logics rooted in American banking and Silicon Valley emerge. First, the alternative lending tools incorporate concepts of creditworthiness that stem from American history and values, wherein one's creditworthiness is not simply about wealth, but also one's character and trustworthiness. This intersects with Silicon Valley logics, in which more data is better, data is the "truth" and leaving decision-making to machine learning algorithms is best given their objectivity.

### 4.1 American banking mindsets

#### 4.1.1 *The purpose of ML-based credit assessment*

Fintechs are grappling with the same question that has been at the forefront for bankers for centuries: will this person repay my loan? As Kamal, an interviewee based in India, put it: "*Ever since money was invented, lending was invented. It's not a new business… But how do you give a loan? How do you evaluate the person?"* The purpose of ML-based credit assessment tools, according to interviewees, is to accurately evaluate whether a person is "creditworthy" to receive a loan, and if so, under what repayment terms. Creditworthiness, captured in the form of a credit score, combines two key aspects: (1) ability to repay, and (2) willingness (or intent) to repay. Kamal continued, "*[There is] sort of this universal law… [For] every bank in the world, every lender in the world, fundamentally there are two things… The ability to repay and the intent to repay*." Across the fintechs, interviewees share that ML-based credit assessment tools are examining and predicting these two aspects to provide a credit score, and subsequent access to loans (or not). ML-based credit assessment tools are thereby assessing one's creditworthiness under the logic that people, even those who are poor, can be creditworthy – as creditworthiness is tied to one's character and willingness to repay as opposed to simply their level of wealth. This echoes the dominant concept and notion of creditworthiness in banking today that has roots in America, and was born under pursuit and promise of meritocracy. There are several ways in which ability to pay and willingness to pay are examined and encoded in the algorithms by fintechs.

#### 4.1.2 *Operationalizing creditworthiness: Ability & Willingness to repay*

Ability to pay is about whether one has the financial means to repay a loan. Ability to pay is largely an assessment of one's job, income, cash flow, and broader financial situation. Kamal noted that estimating ability to repay is, "*fairly simple. How much money is [the person] generally making every month? How much is she spending every month? And how much do you have in the bank? If you are spending a lot more than what you are making, you might be the most honest person in the world, but I can't give you money."* Within this, having an understanding of the stability of one's economic and job situation is important. Fintechs may consider whether one has an informal or formal job, as well as a cyclical/seasonal job or steady job, the latter in both cases being more favorable. Fintechs generally rely on alternative data gathered through smartphones and utilize machine learning to estimate aspects of one's economic and job situation, but may incorporate questionnaires and ask for supporting documentation (e.g., pay slips). Cash flow can be assessed based on how much one is making and spending monthly, which can be tracked over a period of time through assessing one's transactions, withdrawals, and deposits (whether in formal bank accounts, other more informal digital accounts such as MPesa, or through text messages documenting different cash flow metrics). Fintechs may leverage machine learning to estimate cash flows and job stability through a variety of data and variables. Based on the various information and variables, one's ability to pay is assessed. A stable and positive cash flow alongside consistent income would be considered good for the algorithm that is assessing one's ability to pay.

Willingness – or intent – to repay is different. At a high level, this assessment is about whether a person is honest and trustworthy. Willingness to pay could come from various sources, and includes an assessment of one's behavior, particularly one's financial behavior and whether a person is financially disciplined. In addition to financial behavior, it also includes an assessment of the person's reason for taking out a loan (e.g., for education, paying back another loan) and whether one intends to repay or is a fraudster. Willingness to repay is applying numeric understandings to questions of one's honesty, trustworthy, and morality. There are several common approaches in assessing willingness to repay. This can include assessing one's behaviors (e.g., whether one takes out many loans based on how many financial apps they have on their phone, whether one is a gambler based on their phone apps), using psychometric testing to understand one's psychology, and assessing one's relationships with others. If credit scores and credit histories are available on a person in a certain country location, those may be combined with one's smartphone data.

### 4.2 **Silicon Valley mindsets**

There are two beliefs linked to Silicon Valley that are reflected by fintechs and present in ML-based credit assessment tools: the notion that machines are objective and that data is a reflection of truth. These two types of logics, which go hand-in-hand, are pervasive amongst fintechs in the ML credit assessment space and were reflected across the interviewees, including amongst those geographically further from Silicon Valley (e.g., in India).

#### 4.2.1 *"Leave it to the machine"*

Across the interviewees, there is a belief that machine learning tools are objective and less (or no) human involvement or interference is better. This is reflected both in the development of algorithms, as well as trusting in the technology over time. When it comes to developing algorithms for credit assessment, several fintechs discussed disbursing small loans, largely randomly, to see who repays. By distributing loans randomly, fintechs learn from patterns of repayment in that particular time and context, as opposed to learning from past data to make predictions on creditworthiness. This informs what variables are linked to creditworthiness to then develop and inform the algorithms that apply to new applicants. Other fintechs use existing or historical data to inform the model by examining who is "creditworthy" now or historically to develop and inform algorithms that apply to new applicants. In both cases, machine learning identifies patterns in "creditworthy" people to inform algorithms to apply to new applicants. Dante, who is based in the US and works for a fintech that develops ML-based credit assessment models for use globally, noted their reliance on the model to inform what variables are most predictive for assessing credit to then build into a model for new potential borrowers:

> "Depending on the target that we want to analyze, we'll leave it to the machine to really identify the one variable that has the highest information value, the lowest correlation with other features and also the highest stability… Something we know… If a customer has more than 12 apps in the finance category, they tend to be more than two times riskier than somebody with less than five apps in the finance category. Now, how to explain it from a psychological point of view, [a] behavioral psychology point of view? [It] is kind of challenging because we really leave it to the machines to assess informational value… We know also that a customer that takes a disproportionate amount of selfies tends to be a worse repay than somebody that takes very little selfies… You can understand, perhaps the vanity. We don't try to really explain the features in a human intelligible kind of way, but we just provide data statistics."

The idea of "leaving it to the machine" builds from the recognition that humans can be biased. Machines are seen as trusted entities to reduce, or even fully remove, human bias and subjectivity.

Rohan, in the C-suite of a fintech based in India, elucidated: "*We wanted to reduce the subjectivity. Machine learning became one of the more objective methodologies that we could use because it is largely driven by [an] objective set of things… It removes the subjectivity out of this equation. So it makes it a lot more objective in terms of decisions.*" Other interviews echoed that machine learning is objective, and, relatedly, can remove bias. Sarah, based in East Africa who is in a leadership role at a fintech with operations in four country locations, asserted: "*The good thing about the machine learning tool is it also removes any bias because there's no human being who's going through the messages looking at what's right, what's wrong. It just follows a particular path or a particular route.*" These sentiments, echoed across fintechs, reflect a popular notion that technology and AI is objective, while human beings are the ones that are flawed, subjective and/or have bias.

Various interviewees discussed the use of machine learning as a method to remove subjectivity and make lending nondiscriminatory. According to Chris, who is based in Europe and leads a fintech that develops ML-based credit assessment models for use globally: "*The [founder] wanted to create something that was unambiguous. Didn't discriminate, nondiscriminatory, fair. And so the value of just pulling metadata and letting algorithms go to work to find out common features and correlations makes it agnostic to the individual. And so it reduces prejudice from lending and that was a big part… of the thesis and what attracted [venture capital investor] to it… The thing I love about it is absolutely there's no prejudice to it… We're just pulling data from a smartphone and then giving… a score."* These tools are painted as an important tool for avoiding prejudice and discrimination.

#### 4.2.2 *Data is "the truth"*

While machine learning is considered objective, data is seen as the "truth" that feeds the model. These two beliefs go hand-in-hand with each other, as the notion of machine learning being objective relies on data being reflective of fact or truth. Amir, based in India and a co-founder and leader of a fintech operating in India, remarked, "*We don't want to be biased about certain religions, castes, race. We don't ask [about] any of them… We only look at data. And whatever data suggest. I always believe data is the truth. Data will never lie*." This concept of data being the truth does not necessarily consider how data can reflect human choices – for example, where was the data collected from? Who collected it? What data informs the dataset and who is represented amongst that data? In reality, there are various decisions that go into the development of data and selection of datasets. This idea of data being truthful is also central to the belief that these tools are not biased or are less biased than traditional lending. Data is contrasted with the status quo in access to finance, which has been rife with unfair biases stemming from humans. Sarah put it, "*The fact that we eliminated the issue of someone having to show up or fill forms… we don't have to see them. Because there's always this stigma attached to having to walk into a financial institution. Of course, we have our biases. If I look at you, I'll probably assess and say, hmm, this person possibly looks like they can repay or they can't. Even before I look at the data. So the advantage became, I don't need to see you to make a decision. I just need the data to speak to us*." Data is seen as both illustrating and speaking the truth.

### 4.2.3 *Purpose-driven & profit-motivated: Do good & make money*

Many fintechs in the credit assessment space in LMICs, including all those interviewed, are purpose-driven, with missions and goals of reaching people who are underbanked and unbanked with access to finance. They are using technology to enhance access to finance for people underserved by traditional financial institutions and approaches. This sense of purpose is central to the founding and operation of these companies. Josh, a data scientist at a company headquartered in the US with operations in several country locations said, "*This is an awesome opportunity to use devices that people already have to increase their financial agency and autonomy and give them opportunities that they might otherwise not have to get into traditional brick and mortar buildings*." Interviewees shared an excitement for the social impact potential of their organizations and the technologies.

By enabling access to customers otherwise left out of financial folds, machine learning opens new market opportunities that can bring business benefits. As put by Jack, a funder and partner of fintechs: "*From a lender's perspective, it's also good for the bottom line. If you can accurately identify people who are creditworthy that you previously were not lending to, that's a business opportunity."* Lenders are, after all, in the business of making money. At the end of the day, while innovation is prized and purpose is a core driver, fintechs are still companies focused on being sustainable and profitable businesses that continue to attract investment, provide returns, and grow over time. Chris, based in Europe and leading a fintech that provides ML-based credit assessment models to organizations globally, noted: "*Objectivity, realism, execution. That's my mantra. Can we execute well, can we make money doing it*?" Algorithmic-facilitated lending creates a seductive marriage: do good and make money.

Taken together, fintechs carry values and perspectives that are deeply rooted in Silicon Valley culture. Namely that AI technologies are objective and more data is better. Many fintechs emerge from a dual motivation: a commitment to enhancing financial inclusion and a drive for profit. This purpose-driven ethos, common in Silicon Valley, aims to "do good" while simultaneously prioritizing financial returns. The lenses of "technofeminism" and "situated knowledge" emphasize the examination of the social shaping of algorithmic technology.

## 5 FINDINGS B: A GENDER PICTURE EMERGES

Silicon Valley logics of believing in data as the "truth" and "leaving it to the machines" as objective technologies, inform how fintech developers perceive bias and how they operationalize fairness. This perspective often leads fintechs to build models that are "gender blind" or agnostic, viewing this approach as a solution for avoiding bias and achieving fairness. The underlying belief is that making algorithms "blind" to sensitive characteristics and demographics (such as gender and caste) is the best way to ensure impartiality in credit assessments. This means the model does not consider variables of gender (or other demographics) in its credit assessment algorithm.

## 5.1 Gender blind & agnostic

Having algorithms be "blind" to gender was discussed across fintechs as a method to be fair and avoid bias. This was consistently brought up by organizations. Amir noted: "*We try to be unbiased to the core. So, we don't collect gender, we don't collect religion, we don't collect race information*." The idea is to keep the models from knowing demographics of the person and focus on the variables tied to creditworthiness. Suraj, based in California whose fintech has global operations, asserted that being unaware of gender is important for fairness: "*If you think about gender equity, I don't think we take that into consideration. Actually I know we don't take that into consideration, like male or females. That's very fair in that way."* This belief that not considering gender means the model is unbiased to gender is commonplace. Relatedly, interviewees discussed how not considering gender keeps decisions more objective. Sarah, a fintech leader based in East Africa, said: "*Gender doesn't matter in our scoring… Because I'm not using gender in any of our scoring, it means anyone coming onto our platform has a very level playing field with anyone else. You don't have to think about it at the back of your mind, what if?*" Gender is seen as something that could impact judgment of individuals or machines, which would not be desired as it gets in the way of the objectivity of the machines.

This idea of keeping the model objective and avoiding unfair discrimination by not "seeing gender" or other demographics is tied to logics of "leaving it to the machine". Chris captures this connection: "*[We are] completely gender agnostic, color agnostic, race agnostic. This is just a statistical algorithm, machine-based learning that looked at it and then they constantly refined it… It's completely agnostic. We're just pulling data. It brings a real fairness and lack of prejudice to it."* Here, the interviewee almost reduces the role of the fintech and its human employees, implying that it is really the machine and its statistical algorithms that are making decisions and informing the direction. The role of humans is then to acquire the data to feed the machine and then follow the decision from the machine.

## 5.2 Gender can be learned by the algorithms

While fintechs are not incorporating gender into their algorithms, there is acknowledgement that the models can pick up on correlations and learn different aspects of people, including one's gender. The idea that machine learning tools can "learn" demographics has been shown in other examples including in healthcare [53] and hiring [16]. Several interviewees noted that models could pick up on signals to learn gender even if gender is not in their models or explicitly identified. In response to whether gender can be learned by the machine and it informs the machine, Gary, who is based in the US leading a fintech with operations in Asia, responded: "*It definitely gets learned. It definitely learns from that*." How then does gender get learned?

There are different variables and features models can learn from to infer gender. Variables and features can be gendered, resulting in the tools learning gender from those variables and features. In regards to ability to pay, variables such as income and employment status and sector are gendered. Women face higher unemployment rates than men (4.5% globally compared to 4.3% for men) and are more often in informal work, with 80% of new jobs created for women are within the informal economy, whereas for men this number is 66% [52]. Relatedly, up to 92% of working women in low-income countries are in informal employment, versus 87% of men [41]. Income differences are also pronounced, including women being more likely to live in poverty as compared to men. Gender gaps in poverty are highest among those aged 25 to 34, with women in this age range being 1.2 times more likely to live in extreme poverty as compared to male counterparts [48]. These gender gaps are linked to persistent inequities, including – but not limited to – women overrepresented in informal sectors, institutional barriers including workplace discrimination, caretaking norms resulting in women spending more time on unpaid care than men, greater representation in part-time work, gender norms linked to traditional gender roles of men as breadwinners and women as homemakers, and more [48].

Fintechs are aware of and acknowledge how certain variables can be gendered. Kumar explained: "*We don't train them on gender, but then there's a lot of correlated variables when you look at gender… Income is one of them. Like in India, I think [women earn] probably 85 cents for $1 that we [men] earn, right? … But more than that, the sort of employment segment matters a lot*." Another interviewee noted that gender could likely be reverse engineered from the features used by the machine learning model. He used this to question why gender should be incorporated as a feature, because if the model knows gender regardless, there is not a need to incorporate it as its own variable.

Several interviewees acknowledged that the learning of gender may result in the algorithms leaning towards men as borrowers. This is linked to the factors discussed prior, including men being more likely in formal employment and having higher incomes. Prashanth, based in India who is in a leadership role at a fintech with global operations, noted that gender is not incorporated as it is illegal to do so in lending (in the case of India), but *"At a technical level, you know that can still happen, which is not avoidable. It's a preference towards a more stable segment."* Relatedly, Kumar reflected: "*I'm curious about if the tools are more often granting loans to men, which makes sense based on all these different things that exist, right. And just like patriarchy, they [are] kind of learning that men are better to give loans to over time.*" This perception is linked to the aspect of creditworthiness related to *ability to repay* versus *intent*. In particular, it is picking up on gendered differences related to income, job stability, and formal employment that inform 'ability to repay' algorithms. Fintechs do not necessarily think that the gender differences in ML-based credit scoring and loan disbursal are a bad thing. After all, there is a sense they are accurate regarding creditworthiness.

## 5.3 **Enhancing financial inclusion & gender impacts**

### *5.3.1 Financial inclusion overall*

Among the fintechs interviews, the reach of customers varies, with some smaller fintechs reaching tens of thousands of people, while the larger fintechs have reached millions of people – one even claimed to have reached tens of millions of people – including several million monthly active customers. Fintechs are not necessarily tracking how many borrowers facilitated through their algorithms and loan processes are considered "new to credit" or "unbanked". However, several fintechs offered estimated percentages of customers that are “new to credit”. An interviewee in India estimated that 20% of their borrower population would be considered “new to credit”, which equates to about two million people. These numbers can vary based on country location (as credit scoring and formal financial access varies across countries). They also can vary based on the risk appetite of firms and their investors.

Overall, fintechs note how AI and machine learning is very effective for predicting creditworthiness and repayment for unbanked and underbanked people – and continues to get better. The underlying logic is that machine learning tools use data that does not rely on formal access to finance to offer credit scores and loan options. Thereby, fintechs are enabling people who have been outside of financial folds to potentially access loans. This allows them to underwrite people who have not had the opportunity prior. This is affirmed in studies that have found that ML-based credit assessment models in the United States can result in higher rates of credit approvals or lower interest rates for underserved consumers when compared with traditional credit scores [19, 32].

#### *5.3.2 Gender differences in lending*

Several interviewees did not know, or were not willing to share, the gender breakdown amongst users. In response to questions about gender differences in who gets access to loans through the app, Kumar said: “*Generally we haven’t looked into this… I’m not sure whether anyone on the team has. This has never [been] brought up.”* While no interviewees provided raw data of the gender breakdown of borrowers and loan sizes (despite my asking for this data at the end of each interview), over half of the interviewees shared approximate numbers in the interview itself. Others would be vague, noting that more men accessed loans without providing specific numbers. There was a lack of knowing impacts, including as it relates to gender. John noted that they simply don’t have time to track impacts, much less disaggregated by gender.

From the information provided, a gender picture emerges. In short, despite increasing access to finance, this access to finance is not gender equitable, including gender gaps in both the borrowers of loans and loan sizes. Interviewees shared that they provide more loans to men, except in certain countries where the breakdown is more equal. Several interviews discussed loan size, noting that women tend to get smaller loans than men. Although they did not clarify if this is due to being offered smaller loans compared to men or choosing smaller loans even if they were offered higher loans.

Interviewees reported similar gender gaps in borrowing in different country locations. Interviewees that work in multiple country locations note that certain countries have higher gender gaps in algorithmic-facilitated lending (including gender gaps related to borrowers and loan sizes) as compared to others. Josh, who works at a fintech with operations in Kenya, Mexico and the Philippines, said, “*The three different countries are all very different, so they have different ratios of… males and females*.” Algorithmic-facilitated lending tends to have the highest gender gaps in India followed by Kenya and East Africa, compared to other primary market locations (e.g., Mexico, Philippines). In India, Amir acknowledged 20 to 25% of borrowers are female and 75 to 80% are men, while Kumar shared similar numbers at his fintech with 30% of borrowers being female and 70% male. Other interviewees at fintechs in India would not provide specific numbers (whether because they did not know them or because they would not share them). Kamal noted the number of borrowers is "*predominantly men*", while Rohan said simply, "*generally the male goes and applies for a loan. The women typically don't apply*." Sarah acknowledged it as slightly better in Kenya with her fintech having approximately 35% of female borrowers and 65% male borrowers. Interviewees with fintech operations in Mexico and the Philippines shared numbers that were closer to 50% for men and women.

Although a limited sample, these trends track with gender inequality country rankings– meaning that the countries with the larger gender gaps in ML-based lending correlate to countries with higher gender inequality. The Global Gender Gap Index assesses national gender parity and ranks 146 countries. In this, India is listed near the bottom at 129, Kenya in the middle at 75, Mexico near the top at 33, and Philippines nearby at 25 [52]. These trends also track and are linked to gender digital divides, with India having the highest gender digital divide, followed by Kenya, and Mexico and Philippines being closer to parity [33].

#### *5.3.3 Women as better repayers*

An overarching trend is that women are better at repaying the loans. Across the board, interviewees acknowledged women as better repayers, including that they are both more likely to repay on time and to not default. Interviewees acknowledged this as a pervasive trend and offered reasons linked to observations about the trustworthiness and reliability of female borrowers, indicating that this trend remains regardless of loan size. Fintechs acknowledge that from a business perspective, this makes them a better bet. Kamal said: "*Women are proven to be better repayers so from a business point of view it makes sense to actually have women. But that's not what people do*." Sean, who is based in California and is a data scientist leader at a fintech with credit assessment operations largely in sub-Saharan Africa but with intentions to expand more broadly, echoed: "*Women are getting the smaller loans on average, but they are also far better at paying them back. Women are just a much better bet if you are a bank… You want [to lend money] to the women, not the men*." Nicholas, who is based in Europe and is a leader at a fintech with global operations, noted that this is a global trend they have witnessed with women being better repayers in every country location they have worked in. This was further echoed by Dante: "*There are some local rules that apply globally. If you are a woman, you are a better repayer than a man. Full stop.*" This tracks with evidence that women tend to be better at repaying loans, particularly in the context of microfinance [1, 18, 45]. A latent tension exists around stated ideals of opening new roads to finance in meritocratic ways and acknowledgement that women are better repayers. Persistent barriers prevent stated ideals from being met.

### 5.4 **Reasons for gender differences in loans**

#### *5.4.1 Perceptions regarding why there are gender differences*

Interviewees outline several observations regarding why there are gender differences in credit assessment and loans. At a high level, interviewees highlight that gender differences related to loans can be linked to broader gender norms, structural inequities and cultural differences. This manifests in several demand-side factors. First, interviewees note that there tends to be less women applying as compared to men. Some acknowledge that this may be due to less access to apps given the medium of the tool being smartphones and gender differences in mobile ownership. Other interviewees noted that women are also more likely than men to have basic phones, or use a shared phone. Sarah explained: "*There are some communities where you'll find if it is a smartphone and it's one in the household who actually keeps the smartphone and who will keep the feature phone… Being a patriarchal society, if there are two handsets in a home, one more superior than the other, the superior one will belong to the leader of the household, if you will."* Interviewees acknowledge that this factor may be more relevant for rural areas where gender differences in smartphone penetration and Internet tend to be more pronounced. Relatedly, interviewees reflected that less women may choose to apply for loans. Other interviewees noted that because people come to the app, more men generally are those seeking loans as household financial managers. This is captured by Rohan: "*If a person needs a loan it is generally the male who kind of goes and applies for a loan. The women typically don't apply. It's just a social kind of indicator. So wherever it is self-serve and where it's used for consumption, typically you would see the skew towards more men."* Third, even if they do apply, interviewees noted that women can face greater challenges using the app compared to men. Taken together, these factors highlight ways that fintechs have observed demand-side constraints for women in accessing loans.

Looking at the supply side, interviewees reflect on the ways creditworthiness is defined and operationalized in the ML tool. Several discussed how factors related to employment that are considered by the algorithm can have gender implications. Rohan noted: *"I think it's more again a social thing… From an employer workforce in general men are more employed compared to women. There are more women who are, you know, homemakers and other things in India, compared to men. So that's one piece which kind of skews*." Kumar also discussed how predominant roles of women as household caretakers impact credit assessments. He said: "*We don't discriminate against [women], but… Maybe housewives would probably be in a less fair situation to get higher ticket size... They're probably homemakers. So these sorts of things will end up coming in."* This reflection illustrates gendered understandings around features used by algorithms to predict creditworthiness, particularly related to 'ability to pay'.

#### *5.4.2 Applying socio-structural explanations to perceptions of gender differences in loans*

Applying socio-structural explanations to make sense of algorithmic decisions is an approach that can improve the scope of interpretations for model operations. ML models do not operate in isolation, but within complex social and institutional constructs that can significantly impact their behavior and impact [46]. This approach allows me to expose how ML-based credit assessment apps are linked to and embed gender norms and structural inequalities.

The demand- and supply-side factors observed by fintechs are linked to gender norms and structural inequalities. First, fintechs observe that less women are applying to the apps and note that women may have less access to the apps given the medium of the tool being smartphones. Due to gender digital divides less women have access to smartphones and Internet required to access apps. There are persistent global gender differences in digital inclusion and literacy (i.e. the "gender digital divide") as well as financial literacy [29]. Surveys and research by GSMA reveal that across LMICs, women are 8% less likely than men to own a mobile phone and 20% less likely to use the Internet on a mobile, with gender gaps further amplified in rural versus urban areas [44]. In two of the most popular countries for these apps (Kenya and India) gender gaps loom: in Kenya there is a 34% gender gap in mobile Internet use, whereas in India the gender gap is 52% [44]. Unsurprisingly, there is a "fintech gender gap": in a global study of 28 countries, 21% of women use fintech products compared to 29% of men [12].

Gender differences in digital inclusion and literacy, as well as financial literacy link to the other observation by fintechs in which women can face greater challenges using the app compared to men. Linked to persistent gender discrimination and limiting gender norms, women – and those with lower socio-economic status – are more likely to need more support to use financial accounts [51]. While many women have access to the Internet and smartphones, as well as strong digital and financial literacy, the proportion of women with this access and literacy is lower as compared to men, particularly in rural areas.

Secondly, interviewees reflected that less women may choose to apply for loans. This aspect of choice can be linked to gendered household financial decision making. In many countries, bank accounts are more often in the name of the man of the household. This is reflected by gender differences in account ownership globally. Women's account ownership is 6 percentage points lower than men's in LMICs [51], with gender gaps that are further inflated in rural and lower income groups. If loans are distributed to bank accounts, as some fintechs operate, the loan may not be provided to the women. Also, women may not choose to apply or choose lower loan sizes linked to lower tolerance to financial risks. Research illustrates lower levels of tolerance to financial risk given managing family risks and may avoid high risk financial options [15].

Looking at the supply side, the ways creditworthiness is defined and operationalized in the ML tool can favor men inadvertently, linked to gender norms and inequalities in a country. Interviewees acknowledge the gendered nature of certain features or proxies that algorithms consider, particularly regarding algorithms assessing *ability to repay.* These features or proxies – which include income level, consistent or stable income, formal employment, and cash flow – hold and reflect gendered differences and structural inequalities in economic status, employment, and caretaking responsibilities. Women tend to have lower incomes as compared to men, be in more informal employment (versus men as more often in formal employment), have less consistent or stable incomes as compared to men, and tend to have higher caretaking responsibilities or roles as homemakers (versus men as more household financial managers and decision makers) [52]. In assessing cash flow which is a common proxy within algorithms assessing ability to pay, an assessment made on an individual woman's phone may not be reflective of the broader cash flow of her or her family. Rather, if household finances are managed by men in the household, key information could be missing from transactions documented on a woman's phone. Differences in how women and men use their phones, which are linked to gender norms, may also influence the algorithms in other ways that are not fully understood.

Building from the observations and perceptions of fintechs, a picture forms regarding the ways ML-based lending apps are linked to and embed gender norms and structural inequalities. It is important to acknowledge that gender gaps in algorithmic-facilitated lending vary less between fintechs and more between country locations, as aforementioned: countries with greater gender inequality have greater gender gaps in algorithmic-facilitated lending and vice versa.

#### *5.4.3 What is optimized for matters*

There's a key consideration between predicting creditworthiness and optimizing for profitability. Some interviewees shared that ML-based credit assessment algorithms don't just assess risk, they also estimate the "lifetime value" (LTV) of a customer. LTV, a common metric used in lending, reflects expected revenue from interest, fees, and long-term engagement, minus acquisition and servicing costs. Two large fintechs noted that LTV is a key component in their credit assessment. Megan, who is based in California and works at a fintech with global operations, noted: "*You can get to predict lifetime value… And that's how the business can make decisions, like what is our payback window? How much are we worth? Are we willing to wait to get to a net positive return on this user? That's how we make decisions on approval*." By incorporating the potential level of returns over time, assessing creditworthiness begins to reflect more than just one's ability to repay and willingness to repay. Now, factors like how much one may borrow over time, at what interest rate, and with what level of fees or penalties become part of the decision. In this, business priorities, particularly profit optimization, shape decisions. These choices are often not framed as value judgments, but rather good practices for financially sustainable lending. Yet, this framing can obscure how deeply commercial logic is embedded in decisions that affect access to credit even under "for good" umbrellas.

Consideration of profitable loans and borrowers is not new to lending. Indeed, lenders, being in the business of making money, tend to prefer asset wealth clients who can take larger, and thereby more profitable, loans [11]. Gender differences can be present: A report by the International Finance Corporation finds that, among fintechs that do not tailor products to women, only 38% report that women's LTV is higher compared to men's (however, this jumps to 63% for fintechs that customize products and services for women) [31].

In the case of ML-based credit assessment, optimizing algorithms for lifetime value, and thereby profit, may be a contributing factor leaning towards men in credit scoring (who tend to have higher incomes, take higher loans, and reportedly have more late payments that can generate greater returns). This would not be the first documented case of machine learning tools in lending exhibiting a bias due to optimizing for profit: Research on a peer-to-peer lending platform in China found that the introduction of machine learning to inform interest rates resulted in higher interest rates for women. This was not due to a higher estimated risk or lower determined creditworthiness by the algorithm, but because women had lower price sensitivity and the platform could therefore better optimize revenue by offering women loans at higher interest rates [13]. Algorithms optimized for profit may inadvertently penalize women – not because they are riskier, but because they may be less profitable under lifetime value models. In this research, interviews report higher loan sizes and fee-generating behaviors, such as late payments are more often associated with male borrowers. These behaviors have the potential to contribute to higher lifetime value scores and may help explain gendered differences in lending outcomes.

#### *5.4.4 Gender implications & perceived responsibility of fintechs*

While fintechs recognize the role that gender norms and inequities play, they do not thoroughly understand and/or account for gender differences in access and use of phones, as well as ensuing implications for algorithms. They thereby fall prey to "default discrimination". Hannah asserted: "*I mean a lot of it's just the technology is intersecting with the existing norms and culture and society that exists. Then also, you know, maybe for the people building those tools there isn't as much understanding of how it might be different for women.*" No interviewees had an intention to perpetuate gender inequities. Rather it is seen as an unintentional outcome, but is justified because it can still be better than the status quo.

## 6 DISCUSSION

### 6.1 **Algorithms reflect their creators**

In exploring underlying logics that inform technology design and management, the concept of "situated knowledge" reveals how algorithms are results of perspectives and priorities of creators, which are shaped by cultural and institutional contexts and logics. Supporting this analysis, "technofeminism" emphasizes how the social context of technology developers influences design choices. Two primary mindsets and logics rooted in American banking and Silicon Valley emerge.

American banking and Silicon Valley logics influence the conceptualization, design, and management of ML-based credit assessment tools. The alternative lending tools incorporate concepts of creditworthiness that stem from American history and values, wherein one's creditworthiness is not simply about wealth, but also one's character and trustworthiness. The ethos of American banking in how creditworthiness is defined and concepts of credit scoring building from willingness and ability to repay is reinforced and repackaged in ML-based credit scoring tools. Assessment of one's character, morals, and trustworthiness is central – will they *choose* to repay, not just *whether* they can repay. There is an inherent value of meritocracy, in which anyone can lift themselves up with hard work, commitment, and strong character. There is also a subjectivity attached to assessing what is sufficient for one's ability to repay, including preferences for formal versus informal jobs, as well as in assessing one's character. This intersects with Silicon Valley logics, in which more data is better, data is the "truth" and leaving decision-making to machine learning algorithms is best given their objectivity.

The interconnectedness between developers' cultural and institutional contexts and the design of ML tools underscores that knowledge is situated. American banking logics lead to credit scoring algorithms that include assessing one's ability to repay and willingness to repay, which is about someone's morality and trustworthiness. Meanwhile, Silicon Valley logics of believing in data as the "truth" and "leaving it to the machines" as objective technologies, inform choices of fintech developers. In particular, this perspective often leads fintechs to build models that are "gender blind" or agnostic, viewing this approach as a solution for avoiding bias and achieving fairness. This belief in the objectivity of technology and data—coupled with a mission to promote financial inclusion—can result in a limited critical reflection on the subjectivity inherent in the very concept of creditworthiness.

### 6.2 **Belief in the machine's objectivity creates and obscures unintended gender bias**

The concept of "default discrimination" highlights the gap between intention and outcome in technology design. By adopting a 'gender blind' approach, fintechs can obscure existing disparities and fail to directly address the underlying structural inequalities that algorithms then learn from. As a result, ML-based credit assessments may inadvertently perpetuate gender inequities and mask the reproduction of existing power hierarchies.

Gender "blind" or agnostic models result in an inherent flaw: the lack of grappling with the ways data is gendered and reflects inequities, as well as the ways features and proxies connected to ability and willingness to repay are gendered and can be subjective. While there is some recognition that features or proxies used in creditworthiness assessment (e.g., income, stable employment) can be gendered, there is less consensus as to whether this is an issue. The tool is designed to assess creditworthiness, and if these factors are important to creditworthiness, then it is doing its job – even if it is more likely to deem men as more creditworthy. However, in general, there is a lack of consideration on how data can be gendered and lead to misleading or inaccurate predictions regarding creditworthiness. Relatedly, there is a general lack of grappling with gendered differences in regards to access to the apps, which can limit how many women apply to loans via the apps to begin with. This lack of grappling with gender differences in access to apps is not true for every fintech. One fintech in East Africa, for example, has recently started making the credit assessment available through basic phones (in addition to their traditional smartphone app) as part of an effort to increase access to women. However, by and large, there is a lack of grappling with apps being exclusive for many women and not seeing that as an issue the fintech is responsible to solve. This could result in greater data on men versus women that the ML models are trained on.

Data is not necessarily "the truth" and ML models are not objective. What data an ML model is learning from matters, as it can over or under-represent different groups with implications for who it performs better (or worse) for. Data also carries with it a reflection of inequalities that exist in society. Meanwhile, the variables linked to definitions of creditworthiness matter.

Not addressing these gendered aspects of the technology head on is linked to three areas: a deep belief in the objectivity of data and machine learning leading to lack of awareness or recognition this is an issue; avoidance due to regulations around not being able to consider gender and other demographics in scoring and keeping distance by insisting that gender is not seen or taken into account; and/or a sense that it isn't the responsibility of the fintech to address structural gender inequalities. These algorithms are not neutral tools but rather products of specific cultural and institutional frameworks. ML-based credit assessment tools follow Benjamin's 'default discrimination', in which design processes stay "blind" to gender and ignore social cleavages and inequities, thereby defaulting towards continuing discrimination that exists in society.

By applying labels of objectivity to credit scoring via machine learning, power hierarchies and social norms are ignored and inadvertently embedded under veils of objectivity. This highlights an *objective algorithm paradox*, in which the belief in machine learning's objectivity results in being agnostic or 'blind' to certain demographics in ways that hide the gendered nature of algorithms and perpetuate discrimination.

This doesn't necessarily mean that gender should be a factor in algorithmic decision making, but rather auditing across demographics is critical to ensure certain groups aren't being penalized, as well as conducting assessments of how datasets training models and features can be gendered to mitigate issues proactively. Meanwhile, it is important to be transparent about the tradeoffs in algorithmic-facilitated lending and then have discussions, including with impacted consumers and marginalized groups, about what is acceptable or not, as well as how algorithms should be operationalized and considerations around fairness.

### 6.3 **ML-based credit assessment results in gender differences in lending**

While fintechs and their ML-based credit assessment tools are increasing access to finance overall, this access to finance is not gender equitable, as evidenced by gender gaps reported in both the number of loans and loan sizes. Fintechs tend to provide more loans to men and loans at higher amounts. These disparities are more pronounced in countries with higher levels of gender inequality, where gender norms and structural inequities influence both the demand and supply sides of credit. Intersectionality can also play a role in differences.

Interviewees note that gender differences related to loans come from demand-side and supply-side factors, which are linked to gender norms and structural inequality. Fintechs acknowledge that women tend to apply less than men, while also having relatively more challenges in using the app. These differences can be linked to gender digital divides, gender norms around household financial decision-making, and differences in digital and financial literacy. On the supply side, fintechs highlight that features and proxies used in *ability to pay* algorithms in particular (e.g., stable employment, income levels), can be gendered. Gender differences in algorithmic-facilitated lending reported by fintechs correlate to levels of gender inequality, highlighting how ML-based credit assessment apps reflect gender inequalities in the context in which they are deployed. Furthermore, some companies are optimizing for lifetime value of customers, which may be a contributing factor leaning towards men in credit scoring (who tend to have higher incomes, take higher loans, and potentially have more late payments that can generate greater returns).

In all of this, there is a key tension: interviewees observe that women are more likely to repay on time and not default on loans as compared to men, despite being less likely to access loans and have lower loan amounts than men. This tension represents potential discriminatory effects and a market inefficiency resulting from the algorithm, particularly linked to *ability to pay* algorithms. In countries such as Kenya and India where these tools are popular, due to gender norms and structural inequities, women tend to have lower income levels and be employed in formal employment (more often in informal jobs) leading to less consistent income or stable cash flows. While this may be a relevant indicator for creditworthiness for larger loans (e.g., mortgages) these aren't necessarily indicative of ability to pay back smaller loans. An additional factor may contribute to this tension, which is optimizing algorithms along considerations of profit.

The specific mechanisms of gender differences are largely hidden by algorithms that are closed to external evaluation, as well as the "black box" nature of machine learning itself. Even if algorithmic-facilitated lending is more gender equitable than the status quo, this presents the potential of solidifying gender inequities in ways that are not fully understood and projecting them into the future under veils of objectivity. This exploration illustrates how "default discrimination" can manifest in these contexts: fintechs may acknowledge the existence of structural inequalities yet still adopt a "gender blind" approach in algorithmic design, follow other priorities, and ultimately, not tackle inequalities in how they design and manage algorithms.

### 6.4 **Encoded gender norms**

I build from prior theories that examine algorithmic inequality, bias and power broadly and/or in relation to race, to introduce the pattern of *encoded gender norms,* in which the status quo is solidified. In this, there are several key propositions: (1) Data and features are gendered (as is access to technological tools); (2) not considering the ways data and features embedded in algorithms are gendered can replicate and reinforce gender norms; and (3) prioritizing inclusion while also being "gender blind" comes at a cost of equity. This leads to algorithms encoding gender norms in ways that result in self-fulfilling prophecies that become harder to spot and solve. While gender norms can and do evolve in and across societies over time, ML tools may impede this evolution under opaqueness and veils of objectivity. As machine learning tools serve as mirrors, we can expect greater gender variance depending on gender inequality levels, while greater attention will need to be placed in gender unequal areas before potentially risking reinforcing and legitimizing limiting gender norms and inequalities.

## 7 **CONCLUSION**

This paper outlines the underlying logics and mindsets informing the development of algorithmic lending tools (findings A) and how those leading to choices around "gender blind" algorithms and implications for perpetuation of gender inequities in financial access despite women being observed as better repayers (findings B). The idea that technology is inherently objective and, when left alone, will be "fair", is seductive and misleading. In reality, knowledge is situated and algorithms are instruments of values. They reflect the values of the people and – more – the institutions that create them. Believing in the objectivity of machine learning creates and obscures unintended gender bias in ML-based credit assessment. While fintech employees and investors are focused on "doing good", in the context of algorithmic-facilitated lending in LMICs, decisions and tradeoffs ensue. Ironically, that women are better repayers but not getting similar loans through current approaches to algorithmic-facilitated lending illustrates that the promise of meritocracy pursued under American banking logics is not achieved.

## REFERENCES

[1] Beatriz Armendariz de Aghion and Jonathan Morduch. 2010. *The Economics of Microfinance, Second Edition*. The MIT Press. Retrieved August 12, 2024 from https://econpapers.repec.org/bookchap/mtptitles/0262513986.htmhttps://doi.org/10.1038/s41591-024-03113-4

[2] Irani Arraiz, Miriam Bruhn, and Rodolfo Stucchi. 2017. Psychometrics as a Tool to Improve Credit Information. *World Bank Econ. Rev.* 30, Supplement_1 (2017), S67–S76.

[3] Robert Bartlett, Adair Morse, Richard Stanton, and Nancy Wallace. 2019. Consumer-Lending Discrimination in the FinTech Era. https://doi.org/10.3386/w25943

[4] Thorsten Beck, Asli Demirguc-Kunt, and Patrick Honohan. 2009. Access to Financial Services: Measurement, Impact, and Policies. *World Bank Res. Obs.* 24, 1 (2009), 119–145.

[5] Ruha Benjamin. 2019. Race After Technology: Abolitionist Tools for the New Jim Code | Wiley. *Wiley.com*. Retrieved January 17, 2024 from https://www.wiley.com/en-us/Race+After+Technology%3A+Abolitionist+Tools+for+the+New+Jim+Code-p-9781509526437

[6] Tobias Berg, Valentin Burg, Ana Gombović, and Manju Puri. 2020. On the Rise of FinTechs: Credit Scoring Using Digital Footprints. *Rev. Financ. Stud.* 33, 7 (2020), 2845–2897.

[7] Daniel Björkegren, Joshua Blumenstock, Omowunmi Folajimi-Senjobi, Jacqueline Mauro, and Suraj R. Nair. 2022. Instant Loans Can Lift Subjective Well-Being: A Randomized Evaluation of Digital Credit in Nigeria. https://doi.org/10.48550/arXiv.2202.13540

[8] Valentina Brailovskaya, Pascaline Dupas, and Jonathan Robinson. 2021. Is Digital Credit Filling a Hole or Digging a Hole? Evidence from Malawi. Retrieved January 20, 2025 from https://papers.ssrn.com/abstract=3991382

[9] Virginia Braun and Victoria Clarke. 2006. Using thematic analysis in psychology. *Qual. Res. Psychol.* 3, 2 (January 2006), 77–101. https://doi.org/10.1191/1478088706qp063oa

[10] Joy Buolamwini and Timnit Gebru. 2018. Gender Shades: Intersectional Accuracy Disparities in Commercial Gender Classification. In *Proceedings of the 1st Conference on Fairness, Accountability and Transparency*, January 21, 2018. PMLR, 77–91. Retrieved January 20, 2025 from https://proceedings.mlr.press/v81/buolamwini18a.html

[11] Mayra Buvinic and Shubhangi Gokhroo. 2023. A NARRATIVE REVIEW IN LOW- AND MIDDLE-INCOME COUNTRIES. *Cent. Glob. Dev. Work. Pap.* (2023).

[12] S. Chen, S. Doerr, J. Frost, L. Gambacorta, and H. S. Shin. 2023. The fintech gender gap. *J. Financ. Intermediation* 54, (April 2023), 101026. https://doi.org/10.1016/j.jfi.2023.101026

[13] Yongqiang Chu, Chunxing Sun, Bohui Zhang, and Daxuan Zhao. 2023. Fintech and Gender Discrimination. https://doi.org/10.2139/ssrn.4322257

[14] Caroline Criado-Perez. 2021. *Invisible Women: Data Bias in a World Designed for Men|Paperback*. Abrams Press. Retrieved January 20, 2025 from https://www.barnesandnoble.com/w/invisible-women-caroline-criado-perez/1129490658

[15] Rachel Croson and Uri Gneezy. 2009. Gender Differences in Preferences. *J. Econ. Lit.* 47, 2 (June 2009), 448–474. https://doi.org/10.1257/jel.47.2.448

[16] Jeffrey Dastin. 2018. Amazon scraps secret AI recruiting tool that showed bias against women. *Reuters*. Retrieved August 18, 2024 from https://www.reuters.com/article/world/insight-amazon-scraps-secret-ai-recruiting-tool-that-showed-bias-against-women-idUSKCN1MK0AG/

[17] Asli Demirguc-Kunt, Leora Klapper, and Dorothe Singer. 2013. *Financial Inclusion and Legal Discrimination against Women: Evidence from Developing Countries*. World Bank. Retrieved January 20, 2025 from

https://www.researchgate.net/publication/381700880_Financial_Inclusion_and_Legal_Discrimination_against_Women_Evidence_from_Developing_Countries
[18] Bert D'Espallier, Isabelle Guérin, and Roy Mersland. 2011. Women and Repayment in Microfinance: A Global Analysis. *World Dev.* 39, 5 (May 2011), 758–772. https://doi.org/10.1016/j.worlddev.2010.10.008
[19] Marco Di Maggio and Dimuthu Ratnadiwakara. 2022. Invisible Primes: Fintech Lending with Alternative Data. https://doi.org/10.2139/ssrn.3937438
[20] Catherine D'Ignazio and Lauren Klein. 2023. *Data Feminism*. The MIT Press. Retrieved January 20, 2025 from https://mitpress.mit.edu/9780262547185/data-feminism/
[21] Virginia Eubanks. 2018. *Automating Inequality*. St. Martin's Press, New York. Retrieved January 20, 2025 from https://us.macmillan.com/books/9781250074317/automatinginequality/
[22] Federal Reserve Board. 2007. Report to the Congress on Credit Scoring and Its Effects on the Availability and Affordability of Credit. *Federal Reserve*. Retrieved August 12, 2024 from https://www.federalreserve.gov/boarddocs/rptcongress/creditscore/general.htm
[23] Findexable. 2021. *Fintech Diversity Radar: Diversity for Growth*. Retrieved December 7, 2024 from https://findexable.com/fintech-diversity-radar-fdr/
[24] Batya Friedman and Helen Nissenbaum. 2017. Bias in Computer Systems. In *Computer Ethics* (1st ed.). Routledge, 215–232. Retrieved January 20, 2025 from https://www.researchgate.net/publication/329747893_Bias_in_Computer_Systems
[25] FT Partners. 2022. *Women in Fintech*. Retrieved December 7, 2024 from https://www.ftpartners.com/fintech-research/women-in-fintech2022
[26] Grand View Research. 2023. *Global Alternative Financing Market Size & Share Report, 2023-2030*. Grand View Research. Retrieved January 20, 2025 from https://www.grandviewresearch.com/industry-analysis/alternative-financing-market-report
[27] Jarek Gryz and Marcin Rojszczak. 2021. Black box algorithms and the rights of individuals: no easy solution to the "explainability" problem. *Internet Policy Rev.* 10, 2 (June 2021). Retrieved August 18, 2024 from https://policyreview.info/articles/analysis/black-box-algorithms-and-rights-individuals-no-easy-solution-explainability
[28] Donna Haraway. 1988. Situated Knowledges: The Science Question in Feminism and the Privilege of Partial Perspective. *Fem. Stud.* 14, 3 (1988), 575–599. https://doi.org/10.2307/3178066
[29] Andrea Hasler and Annamaria Lusardi. 2017. *The Gender Gap in Financial Literacy: A Global Perspective*. George Washington University, Global Financial Literacy Excellence Center. Retrieved January 20, 2025 from https://gflec.org/research/
[30] InsightAce Analytic. 2024. *AI In The Credit-Scoring Market Latest Trends Analysis Report in 2024*. InsightAce Analytic. Retrieved January 20, 2025 from https://www.insightaceanalytic.com/report/ai-in-the-credit-scoring-market/2578?utm_source=whatech&utm_medium=refferal&utm_campaign=shorturl&utm_content=whatech-com-902002
[31] International Finance Corporation. 2024. Her Fintech Edge: Market Insights for Inclusive Growth. *IFC*. Retrieved April 8, 2025 from https://www.ifc.org/en/insights-reports/2024/her-fintech-edge-market-insights-for-inclusive-growth
[32] Julapa Jagtiani and Catharine Lemieux. 2019. The roles of alternative data and machine learning in fintech lending: Evidence from the LendingClub consumer platform. *Financ. Manag.* 48, 4 (2019), 1009–1029. https://doi.org/10.1111/fima.12295
[33] Ridhi Kashyap and Ingmar Weber. 2023. Digital Gender Gaps. *University of Oxford*. https://doi.org/10.5281/ZENODO.7897491
[34] Thomas S. Kuhn. 1970. *The structure of scientific revolutions* ([2d ed., enl ed.). University of Chicago Press, Chicago.

[35] Josh Lauer. 2017. *Creditworthy: A History of Consumer Surveillance and Financial Identity in America*. Columbia University Press.

[36] Donald MacKenzie and Judy Wajcman. 1999. *The Social Shaping of Technology*. Oxford University Press. Retrieved January 17, 2024 from https://www.research.ed.ac.uk/en/publications/the-social-shaping-of-technology-3

[37] David Mhlanga. 2021. Financial Inclusion in Emerging Economies: The Application of Machine Learning and Artificial Intelligence in Credit Risk Assessment. *Int. J. Financ. Stud.* 9, 3 (September 2021), 39. https://doi.org/10.3390/ijfs9030039

[38] MicroSave. 2019. Making digital credit truly responsible- Insights from Kenya. *MicroSave Consulting (MSC)*. Retrieved January 20, 2025 from https://www.microsave.net/2019/09/18/11310/

[39] Safiya Umoja Noble. 2018. *Algorithms of Oppression: How Search Engines Reinforce Racism*. NYU Press. https://doi.org/10.2307/j.ctt1pwt9w5

[40] Ziad Obermeyer, Brian Powers, Christine Vogeli, and Sendhil Mullainathan. 2019. Dissecting racial bias in an algorithm used to manage the health of populations. *Science* 366, 6464 (October 2019), 447–453. https://doi.org/10.1126/science.aax2342

[41] OECD and ILO. 2019. *Addressing the gender dimension of informality*. OECD, Paris. https://doi.org/10.1787/cfd32100-en

[42] María Óskarsdóttir, Cristián Bravo, Carlos Sarraute, Bart Baesens, and Jan Vanthienen. 2020. Credit Scoring for Good: Enhancing Financial Inclusion with Smartphone-Based Microlending. https://doi.org/10.48550/arXiv.2001.10994

[43] Jonathan Robinson, David Sungho Park, and Joshua Evan Blumenstock. 2023. The Impact of Digital Credit in Developing Economies: A Review of Recent Evidence. https://doi.org/10.2139/ssrn.4540063

[44] Oliver Rowntree, Kalvin Bahia, and Caroline Butler. 2020. *The Mobile Gender Gap Report 2020*. GSMA. Retrieved January 20, 2025 from https://data.gsmaintelligence.com/research/research/research-2020/the-mobile-gender-gap-report-2020

[45] Abu Zafar M. Shahriar, Luisa A. Unda, and Quamrul Alam. 2020. Gender differences in the repayment of microcredit: The mediating role of trustworthiness. *J. Bank. Finance* 110, (January 2020), 105685. https://doi.org/10.1016/j.jbankfin.2019.105685

[46] Andrew Smart and Atoosa Kasirzadeh. 2024. Beyond Model Interpretability: Socio-Structural Explanations in Machine Learning. https://doi.org/10.48550/arXiv.2409.03632

[47] Tavneet Suri, Prashant Bharadwaj, and William Jack. 2021. Fintech and household resilience to shocks: Evidence from digital loans in Kenya. *J. Dev. Econ.* 153, (November 2021), 102697. https://doi.org/10.1016/j.jdeveco.2021.102697

[48] UN Women. 2023. Progress on the Sustainable Development Goals: The gender snapshot 2023. *UN Women*. Retrieved July 22, 2024 from https://www.unwomen.org/en/digital-library/publications/2023/09/progress-on-the-sustainable-development-goals-the-gender-snapshot-2023

[49] Judy Wajcman. 2006. *TechnoFeminism* (repr ed.). Polity, Cambridge.

[50] Sarah Myers West. 2019. Discriminating Systems: Gender, Race, and Power in AI - Report. *AI Now Institute*. Retrieved October 16, 2024 from https://ainowinstitute.org/publication/discriminating-systems-gender-race-and-power-in-ai-2

[51] World Bank. 2021. The Global Findex Database 2021. *World Bank*. Retrieved August 18, 2024 from https://www.worldbank.org/en/publication/globalfindex/Report

[52] World Economic Forum. 2024. Global Gender Gap Report 2024. *World Economic Forum*. Retrieved July 23, 2024 from https://www.weforum.org/publications/global-gender-gap-report-2024/

[53] Yuzhe Yang, Haoran Zhang, Judy W. Gichoya, Dina Katabi, and Marzyeh Ghassemi. 2024. The limits of fair medical imaging AI in real-world generalization. *Nat. Med.* (June 2024), 1–11. https://doi.org/10.1038/s41591-024-03113-4